\documentclass[sigconf]{acmart}

\setcopyright{rightsretained}

\usepackage{soul}

\usepackage[utf8]{inputenc}
\usepackage{listings} %For code 
\usepackage[linesnumbered,ruled,vlined]{algorithm2e}
\usepackage{graphicx} %Loading the package
\usepackage{subfigure}
\graphicspath{{images/}} %Setting the graphicspath

\newcommand{\toolName}{SLGPT}
\newcommand{\scaleFactor}{0.5} % for images
\newcommand{\MSF}{model file}
\newcommand{\MSFUnit}{line} %% CC says: sometimes you need singular

\makeatletter
\@ifclasswith{acmart}{manuscript}
{\newcommand{\minipageDis}{0.3}}{\newcommand{\minipageDis}{0.2}}
\makeatother

\begin{document}

\author{Sohil Lal Shrestha}
\affiliation{%
  \department{Computer Science and Engineering Department}
  \institution{University of Texas at Arlington}
  \streetaddress{500 UTA Blvd}
  \city{Arlington}
  \state{Texas}
  \postcode{76019}
  \country{USA}
}

\author{Christoph Csallner}
\affiliation{%
  \department{Computer Science and Engineering Department}
  \institution{University of Texas at Arlington}
  \streetaddress{500 UTA Blvd}
  \city{Arlington}
  \state{Texas}
  \postcode{76019}
  \country{USA}
}

\title[SLGPT: Using Transfer Learning to Find Simulink Toolchain Bugs]{\toolName{}: Using Transfer Learning to Directly Generate Simulink Model Files and Find Bugs in the Simulink Toolchain}

\begin{abstract}
Finding bugs in a commercial cyber-physical system (CPS) development tool such as Simulink is hard as its codebase contains millions of lines of code and complete formal language specifications are not available. While deep learning techniques promise to learn such language specifications from sample models, deep learning needs a large number of training data to work well.  \toolName{} addresses this problem by using transfer learning to leverage the powerful Generative Pre-trained Transformer 2 (GPT-2) model, which has been pre-trained on a large set of training data. \toolName{} adapts GPT-2 to Simulink with both randomly generated models and models mined from open-source repositories. \toolName{} produced Simulink models that are both more similar to open-source models than its closest competitor, DeepFuzzSL, and found a super-set of the Simulink development toolchain bugs found by DeepFuzzSL.
\end{abstract}

\copyrightyear{2021}
\acmYear{2021}
\acmConference[EASE 2021]{Evaluation and Assessment in Software Engineering}{June 21--23, 2021}{Trondheim, Norway}
\acmBooktitle{Evaluation and Assessment in Software Engineering (EASE 2021), June 21--23, 2021, Trondheim, Norway}\acmDOI{10.1145/3463274.3463806}
\acmISBN{978-1-4503-9053-8/21/06}




\maketitle

\section{Introduction}

Finding bugs in a commercial cyber-physical system (CPS) development tool such as Simulink is hard as its codebase contains millions of lines of code and complete formal language specifications are not available. While deep learning techniques promise to learn such language specifications from sample models, deep learning needs a large number of training data to work well and the closest related deep learning tool DeepFuzzSL~\cite{Shrestha20DeepFuzzSL} is severely limited by the relatively small number of available training models.

Testing CPS development tools is important as engineers design and develop dynamic safety-critical systems using these development tools. For example, MathWorks's Simulink is widely used in industry such as automotive, medical, and aerospace~\cite{CPS:Zheng}. Engineers use Simulink to design, simulate, test, and generate embedded code from CPS models and deploy it to end-user hardware. At worst a subtle bug in the Simulink tool chain could result in unexpected behaviour in safety-critical applications such as in cars or airplanes.

Given the complexity of the Simulink language, training a deep learning tool such as DeepFuzzSL from scratch would require a very large number of training models. However relatively few open source Simulink models are available. While random model generators such as SLforge~\cite{chowdhury2018icse} could fill in some of these gaps, it is not clear how well SLforge can cover the various features (and their combinations) of the Simulink language.

Given the limited amount of Simulink training models, this paper proposes to use transfer learning for generating Simulink models. Transfer learning is a promising alternative to learning from scratch, as it leverages a machine learning model trained on a large set of related training data. We can then use a relatively small set of Simulink-specific training data to fine-tune such a pre-trained model for generating Simulink models.

Here, we fine-tune the Generative Pre-trained Transformer~2 (GPT-2)~\cite{nlp:gpt2radford2019language} model using both randomly generated models and models we mined from the open-source repositories GitHub and MATLAB Central.
Our experimental results suggest that GPT-2 generated Simulink models are of higher quality and address the shortcomings of earlier
deep learning
approaches. \toolName{} also found a wider range of similar bugs found by DeepFuzzSL in Simulink versions R2018b, R2019b, and R2020b confirmed by Mathworks Support.
To summarize, the paper makes the following contributions.

\begin{itemize}
    \item \toolName{} is the first use of transfer learning for generating graphical block-diagram models. 
    
    \item The paper implements \toolName{} for Simulink, collects a training set of 400 open-source Simulink models, and compares \toolName{} with the closest related tool DeepFuzzSL.

    \item \toolName{}-created models were more similar to open-source models and \toolName{} found a super-set of the Simulink development 
    toolchain bugs DeepFuzzSL found.
    
    \item The \toolName{} implementation, parameter settings, and training sets are open-source~\cite{code:SLGPT}.

\end{itemize}

\section{Background}

\begin{figure}
    \centering
    \begin{minipage}{\minipageDis\textwidth}
        \includegraphics[scale =0.8]{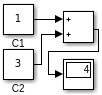}
    \end{minipage}
    \begin{minipage}{\minipageDis\textwidth}
        \includegraphics[scale =0.35]{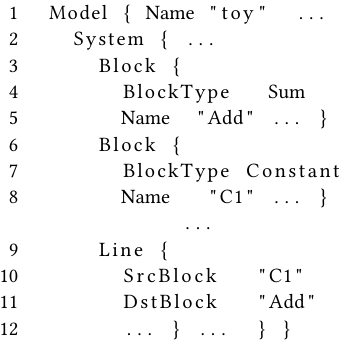}
    \end{minipage}
    \caption{Simulink model (left) and excerpt of its 1.1k~\MSFUnit{} (Simulink-generated) \MSF{} representation (right).
    }
    \label{fig:original_file}
\end{figure}

Simulink~\cite{doc:simulink} is a powerful commercial tool-chain for model-based design and has become a de-facto standard in several domains such as automotive and aerospace. An engineer typically designs a
model via Simulink's graphical modeling environment. A Simulink model is a (potentially hierarchical) \emph{block diagram}, where each block represents equations or modeling components. A Simulink user can also define \textit{custom blocks} in custom ``native'' code using the S-function interface. Simulink typically stores a model in its proprietary \MSF{} format, i.e., as a structured ASCII file that contains keywords and parameter-value pairs (many of which are case-sensitive)~\cite{doc:R2007b}. Figure~\ref{fig:original_file} shows a flat Simulink model and parts of its \MSF{} representation. 

Depending on the block type, each block can accept input via input ports, perform some operation on its inputs, and pass output via output ports to other blocks through (directed) edges. Simulink users can add blocks from various built-in \textit{libraries} and toolboxes. A source block generates signals in a Simulink model while a sink block is used to display or output signals\cite{doc:blocklib}. A model's maximum source-to-sink path length is the longest directed non-circular path from a source to a sink node (and includes source and sink).

When a user opens a model, Simulink's parser performs its checks and prevents corrupt models from opening. Once opened, a user can compile and then simulate the model, where the tool chain uses configurable solvers to iteratively solve the model's network of mathematical relations via numerical methods, yielding for each output block a sequence of outputs. After simulation, the user may use Simulink's embedded code generation workflow for deployment on a target platform.

\subsection{Transfer Learning \& NLP Language Models} \label{sec:lang}

\emph{Transfer learning}~\cite{others:TransferLearningSurvey} is a promising technique for generating Simulink models, as transfer learning can work well in scenarios that suffer from relatively small amounts of training data. Transfer learning achieves this by using a machine learning model trained for a source task or domain (``pre-training'', e.g.,  programs in any programming language) as a starting point to train on a different but related target task or domain (``fine-tuning'', i.e., Simulink models). This works well if pre-training uses huge amounts of training samples, learns features common to both tasks, and fine-tuning can apply the learned knowledge on a target task. Successful applications include computer vision, where large datasets such as ImageNet~\cite{others:imagenet} have been used to pre-train deep learning models that are later fine-tuned for  
tasks such as image segmentation. 

In natural language processing (NLP), language modeling is the use of statistical techniques to determine the probability of a given word sequence. A language model basically estimates the probability of a word based on the words already observed in a sequence. An effective language model not only understands language structure (syntax) but also long-term context (semantics). For example, a Simulink language model should predict tokens that are both syntactically correct and produce valid connections between blocks (e.g., respecting Simulink language rules on define-before-use).

Transfer learning in natural language processing is relatively new. ULMFiT presents a specific training schedule enabling transfer learning using LSTMs~\cite{nlp:ULMfit}. GPT-2 uses transformer decoder as a building block and trains a language model on the WebText dataset~\cite{nlp:gpt2radford2019language}. Using transformers instead of LSTMs allows longer-range context capture. GPT-2's byte pair encoded vocabulary also supports Unicode (and does not require common pre-processing steps such as lower-casing and stemming). So GPT-2 is a great candidate to learn Simulink model files.

\section{Overview and Design}

\begin{figure}
    \centering
    \includegraphics[scale=0.45]{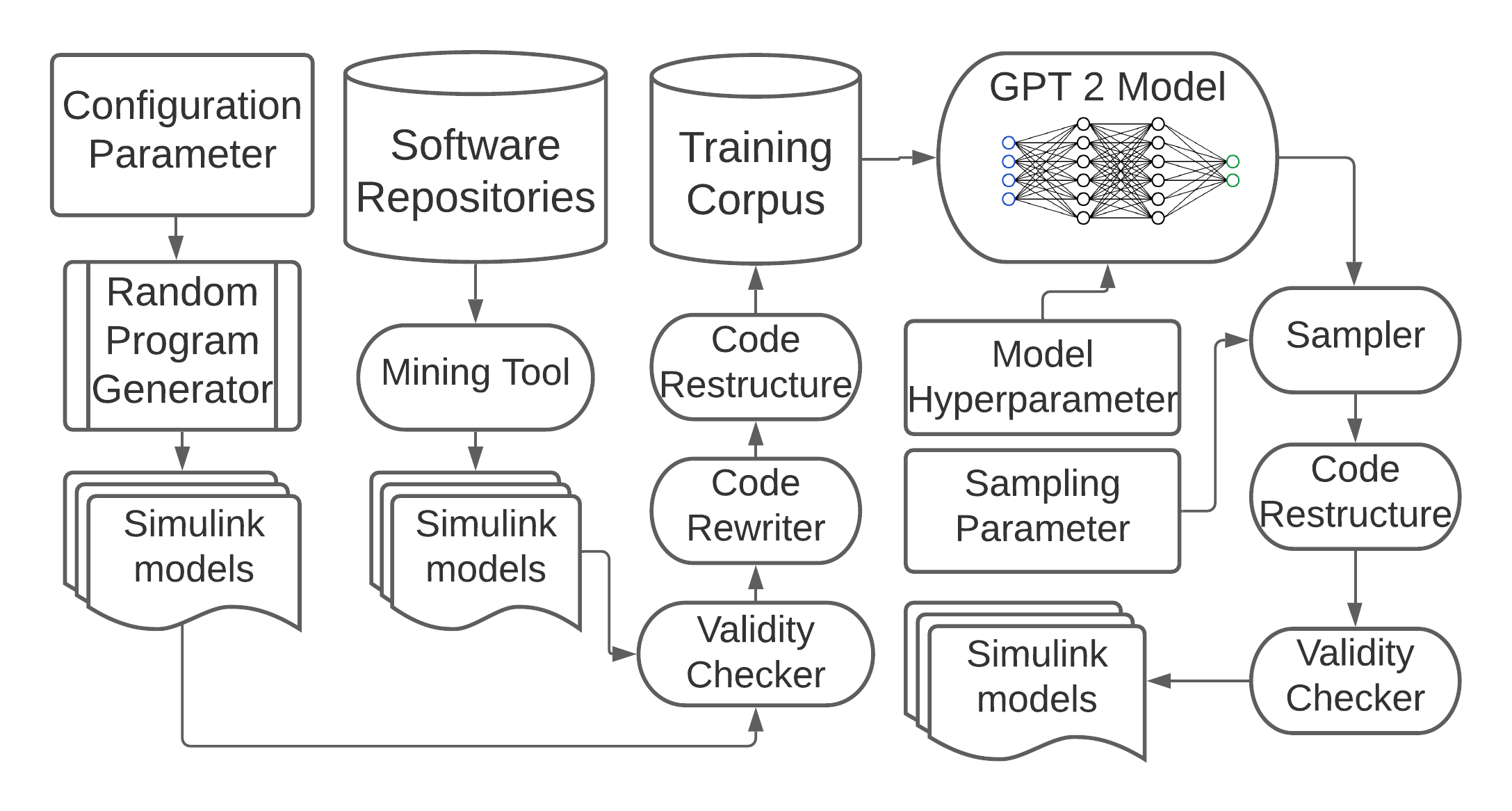}
    \caption{\toolName{} obtains Simulink models from a random generator and open-source repositories, simplifies them, and uses them to adapt GPT-2 for finding Simulink crashes.}
    \label{fig:overview}
\end{figure}

Figure~\ref{fig:overview} gives an overview of \toolName{}.
To obtain a variety of Simulink models for machine learning, we both ran the random model generator SLforge and mined open-source repository sites, i.e., GitHub and MATLAB Central. Since GitHub currently does not treat Simulink as a searchable language, we used the GitHub API with "Simulink" as search keyword. Since MATLAB Central does not provide an API for downloading Simulink models, we used its RSS feed\footnote{https://www.mathworks.com/company/rss.html} to heuristically construct Simulink project download links.

We want our training corpus to only contain valid Simulink models. So we automate the process of checking if a Simulink model is compilable on Simulink. The validity checker also helps detect any crashes caused by an input Simulink model, which is then manually reviewed and reported to the developers. 
To limit the number of Simulink language features in our training data, we only used flat models that do not have additional toolbox or library dependencies, yielding 400 valid open-source Simulink models for training.

\subsection{Training Data Preparation: Simplification}

\toolName{} simplifies training models to (1)~remove model features we currently cannot handle given the limited number of training models and to (2)~restructure models to fit GPT-2's learning style. While both simplification types may change model semantics, \toolName{} compensates for type-2 simplifications (restructuring), by rewriting generated models into equivalent Simulink-compliant style.

\label{sec:corpusprep}

\begin{figure}
    \centering
        \begin{minipage}{\minipageDis\textwidth}
            \includegraphics[scale =0.8]{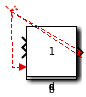}
        \end{minipage}
        \begin{minipage}{\minipageDis\textwidth}
           \includegraphics[scale =0.35]{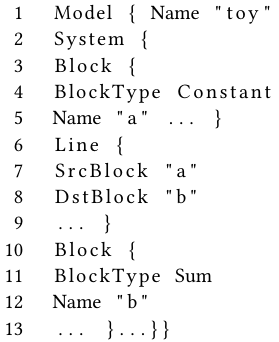} 
        \end{minipage}
    \caption{Figure~\ref{fig:original_file} Simulink model and excerpt of its \MSF{}, after \toolName{} simplified it to 45 \MSFUnit{}s, by removing layout info, restructuring code via BFS, etc.}
    \label{fig:restructed_file}
\end{figure}
\begin{figure}
    \centering
    \begin{minipage}{\minipageDis\textwidth}
        \includegraphics[scale =0.8]{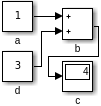}
    \end{minipage}
    \begin{minipage}{\minipageDis\textwidth}
    \centering
        \includegraphics[scale =0.35]{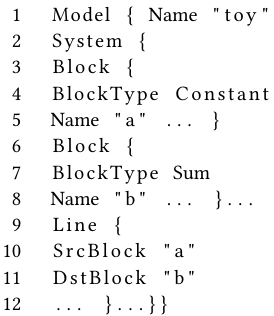}
    \end{minipage}
     \caption{Figure~\ref{fig:restructed_file} Simulink model and excerpt of its 45 \MSFUnit{} \MSF{}, after \toolName{} restored it to Simulink-compliant
     style (plus manual layout changes for readability).
     }
    \label{fig:reconstructed_file}
\end{figure}

Specifically, we pre-process the \MSF{} to remove macros, 
default configuration settings, comments, duplicate white spaces, annotations, and block-position information. We similarly rewrite model identifiers (e.g., block names) to short but unique names (a, b, c, \dots, aa, ab, ac, \dots), based on their appearance order in our restructured \MSF{}.

\label{sec:coderestructured}

The ASCII style in which Simulink saves its models to files is problematic for state-of-the-art deep learning language models, as Simulink files are long and verbose. Furthermore, these files also list all nodes before all edges. Taken together, this is a poor fit for current language models, which model context with a text window of limited size.

\begin{algorithm}[h]
  \caption[caption]{Restructuring Simulink model. ``Neighbour'' refers to both incoming and outgoing blocks and edges.\footnotemark{}}

   \label{algo:bfsrestructure}
\SetKwInput{KwData}{Require}
\SetKwInput{KwResult}{Result}
\SetAlgoLined
\KwData{$source\_blks$ $(S)$, $other\_blks$ $(B)$, $graph\_info$ $(G)$\\  }
\KwResult{BFS-rewrite of Simulink \MSF{} ($C_{BFS}$)}
 \While{S $\neq \emptyset$ \st{and} or  $B \neq \emptyset$}{
    Q = empty queue \\
    b = remove element from $(S\neq\emptyset) ? S : B$ \\
    add b to back of Q\\
      \While{Q $\neq \emptyset$}{
          $curb$ = pop element from front of $Q$\\
          \If{$curb \notin C_{BFS}$}{
              add $curb$ to $C_{BFS}$ \\
              remove $curb$ from $B$ \\
              $B_{nei}, E_{nei}$ = $curb$'s neighbour blocks, edges in $G$ \\
              \ForAll{$e \in E_{nei}$}{
                \If{ $e \notin$ $C_{BFS}$}{
                    add $e$ to $C_{BFS}$
                }                
              }
              \ForAll{$b \in B_{nei}$}{
                \If{ $b \notin$ $C_{BFS}$}{
                    add $b$ to back of $Q$
                }
             }
         }
    }
  }
\end{algorithm}
  \footnotetext{Changes from published version: In Algorithm 1, while condition in line 1 changed from ``and`` to ``or``}

To make such files easier to learn, \toolName{}'s Algorithm-\ref{algo:bfsrestructure} rewrites these files in a breath-first search (BFS) style. Specifically, we first parse the Simulink \MSF{} and maintain an adjacency representation of the Simulink model in the $graph\_info$ map, which maintains two disjoint sets: $source\_blks$ has blocks with $in\_degree = 0$ and other $other\_blks$ has all remaining blocks.
Algorithm-\ref{algo:bfsrestructure}'s outer loop iterates over both $S$ and $B$ as some Simulink models have dangling blocks (blocks that are not source blocks and are not connected to any other block). Also some models (especially from SLforge) have no source blocks because they have cycles.

\subsection{Synthesizing Simulink Models with GPT-2}

Given the complexity of the Simulink language, generating valid Simulink model files is an ambitious task for unsupervised machine learning, especially given our small amounts of training data. Instead of training from scratch, we thus use the pre-trained language model GPT-2. GPT-2 is a good fit, as it employs byte pair encoding to construct its vocabulary, meaning all tokens in a Simulink \MSF{} can be mapped to the vocabulary set.

Second, GPT-2's architecture is based on the transformer architecture~\cite{nlp:transformer}, which has benefits over a traditional LSTM architecture, as transformers avoid recursive computation by processing sequences as a whole and learning relationships between tokens by using multi-head attention mechanisms and positional embeddings. This enables better prediction, which is typically lost with LSTM over long-term dependencies in the text.

\toolName{}'s Algorithm-\ref{algo:sampling} iteratively samples from the fine-tuned language model to generate Simulink \MSF{}s. We seed the model with the sequence ``Model \{'' and then sample token by token. In this early project stage we followed the best sampling techniques of DeepFuzzSL (nucleus or top-p sampling~\cite{nlp:samplingtechniques}). Specifically, given a start text $S$, sampling parameters nucleus $N$ and temperature $T$, the fine-tuned GPT-2 model $G$ computes the probability mass function $PMF$ representing the probability distribution of all tokens in the vocabulary. We normalize the $PMF$ after scaling with $T$ to introduce randomness. To reduce the size of next plausible tokens, we select the smallest subset of $PMF$ such that the sum of all values in the subset is greater than $N$. The normalized subset $PMF$ is then used to perform a multinomial experiment to choose the next token.  

\begin{algorithm}[h]
  \caption{Sampling a candidate Simulink model from a seed text.}
   \label{algo:sampling}
\SetKwInput{KwData}{Require}
\SetKwInput{KwResult}{Result}
\SetAlgoLined
\KwData{Fine-tuned GPT-2 model $(G)$, temperature $(T)$, nucleus $(N)$}
\KwResult{Completed sample string S}
        $S$ = "Model \{ " \\
        \While{<endoftext> $\notin S$} {
            PMF = get\_distribution\_of\_next\_predicted\_tokens$(G,S)$ \\
            Scale the obtained PMF by $T$ \\
            Sort PMF in descending order\\
            Subset PMF such that the smallest possible set sum is greater than $N$ \\
            $R =$ Perform multinomial experiment on subset PMF \\
            $S = S+R$ \\
        }
\end{algorithm}

Since the resulting Simulink \MSF{} $S$ is (as the training samples) in BFS style (as Simulink expects block definitions before edge definitions in a \MSF{}),

\toolName{} restructures $S$ such that the model defines all blocks before defining edges. To continue the Figure~\ref{fig:original_file} example, if we assume Figure~\ref{fig:restructed_file} shows a model produced by Algorithm-\ref{algo:sampling}, \toolName{} then reorders its element definitions to the Simulink-friendly style of Figure~\ref{fig:reconstructed_file}.

In lieu of full differential testing, \toolName{} just uses its validity checker 
to detect crashes of the Simulink tool. We then manually investigate each crash, judge if a crash is an example of a known bug, and report representatives of the remaining crashes to MathWorks Customer Support.

\section{Initial Experience}

While a full evaluation is future work, this paper
compares \toolName{} to its most closely related competitor, i.e., DeepFuzzSL.

We first used DeepFuzzSL's evaluation setup of a SLforge-generated training corpus, in which each Simulink model has 5--57 blocks. \toolName{}'s pre-processing reduced the number of tokens by 75\%, yielding 987 Simulink models with a total of 0.5M \MSFUnit{}s. We ran a related experiment on the 400 open-source Simulink models. 

\toolName{}'s pre-processing removed the 23 of the 400 models that only contained annotation blocks, yielding 3.5k blocks represented in 87k \MSFUnit{}s.

OpenAI has released four different sizes of pre-trained GPT-2 models ranging from 0.1 to 1.5 billion parameters.
To limit computational resource needs, for these initial experiments we used the smallest model.
We fine-tuned the GPT-2 model remotely in the high performance Texas Advanced Computing Center (TACC)'s Maverick~2 cluster~\cite{tacc:homepage}. We ran our experiments on a single Maverick~2 GTX node\footnote{https://portal.tacc.utexas.edu/user-guides/maverick2} 
of two 8-core 2.1~Ghz Intel Xeon processors, 128 GB RAM, and 4~NVidia 1080-TI GPUs. 

As in DeepFuzzSL's experimental setup we used the Adam optimizer (here to fine-tune the GPT-2 model). We could not use a mini batch size of 64 as it triggered out-of-memory errors on TACC. Instead, we used batch size of 1. To compensate for the low batch size, we set the learning rate to 0.00002 (vs. 0.002) and trained \toolName{} for 24 hours on SLforge-generated models and in a separate experiment for 24 hours on the open-source models.

We trained DeepFuzzSL on the same hardware as \toolName{} but otherwise as described in its paper, i.e., for 400 epochs with mini batch size 64. While this ``only'' took about 6.5 hours, DeepFuzzSL's loss function tapered off after 100--150 epochs (so it was not learning much after that).
While sampling, we let DeepFuzzSL run until it either emits a terminating token or reached 15k tokens (corresponding to the largest open-source training model). In the latter case the resulting file typically contained several model-start sequences. When opening such a file, Simulink and our counts just ignore all but the first model.
We use the following research questions.
\begin{description}
    \item[RQ1]  Can \toolName{} generate valid Simulink models? How does the structure of DeepFuzzSL and \toolName{} generated Simulink models compare to open-source models?
    \item[RQ2] How do DeepFuzzSL and \toolName{} compare in the bugs they find in the Simulink tool chain?
\end{description}

\subsection{\toolName{} Can Generate Valid and More Realistic Simulink Models (RQ1)}

Earlier approaches were evaluated in terms of the validity of generated models and their bug-finding ability (e.g., in SLForge, SLEMI, and DeepFuzzSL~\cite{chowdhury2018icse,Chowdhury20SLEMI,Shrestha20DeepFuzzSL}). In addition, to evaluate the quality of a model generator, we compare structural properties of the generated Simulink models against open-source Simulink models. Specifically, we use the number of nodes in the generated Simulink model and metrics based on the common notion of a connected subgraph (i.e., a subgraph in which each node is connected to at least one other node in the subgraph).

To explore \toolName{}'s ability to generate valid Simulink models, we continuously generate Simulink models for 24 hours. Sampling the version trained on SLforge-generated models yielded 2,912 Simulink models of which 43\% compiled. The version trained on open-source models yielded 709 Simulink models of which 47\% compiled. 
The most frequent cause of compile errors include data type mismatches between two connecting blocks and assigning an alphanumeric value to a numeric block attribute. Most of these could be fixed easily by adding data type conversion blocks to the model and changing alphanumeric to numeric values.

We trained DeepFuzzSL on the same training sets as \toolName{} and sampled around 1k samples each with DeepFuzzSL's sampling heuristics. To make the comparison consistent we removed DeepFuzzSL's output token bound and allowed DeepFuzzSL to generate complete Simulink \MSF{}s. Of around 1,200 DeepFuzzSL-generated models trained on Slforge-generated models, 89\% compiled, closely aligning with the 90\% validity rate reported in the DeepFuzzSL paper. On the other hand,   out of 1,024 DeepFuzzSL-generated Simulink models trained on open-source models only 42\% compiled. 

\begin{figure*}[h]
    \includegraphics[width=.24\textwidth]{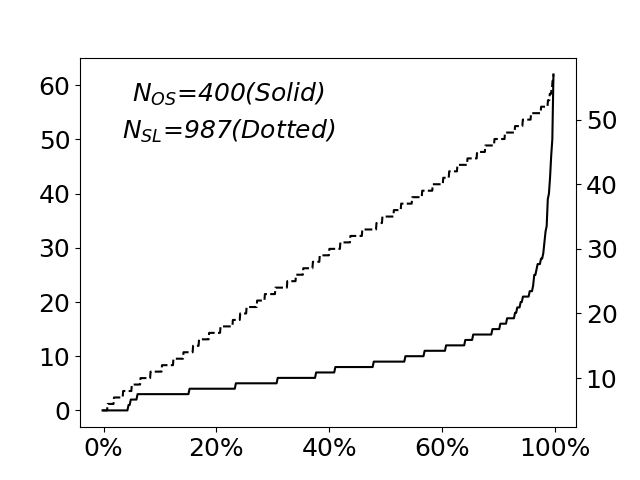}\hfill
   \includegraphics[width=.24\textwidth]{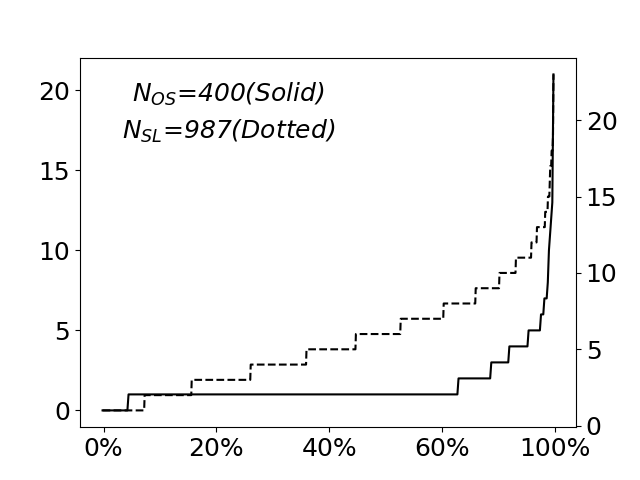}\hfill
\includegraphics[width=.24\textwidth]{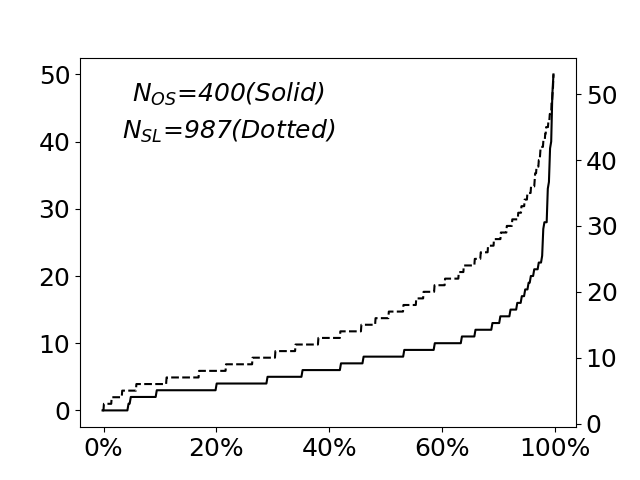}\hfill
   \includegraphics[width=.24\textwidth]{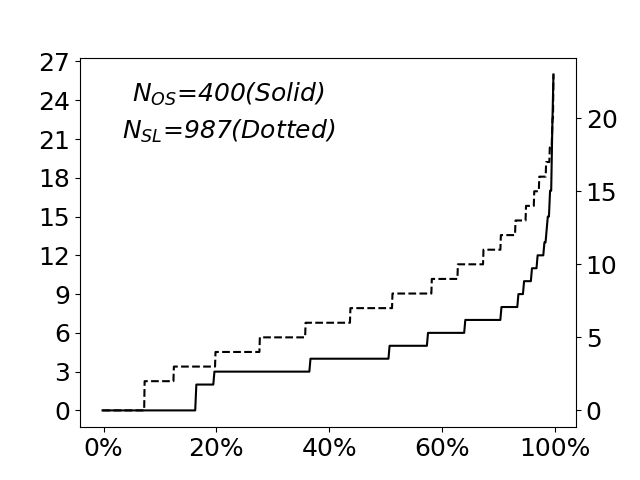}\hfill
        \\
         \includegraphics[width=.24\textwidth]{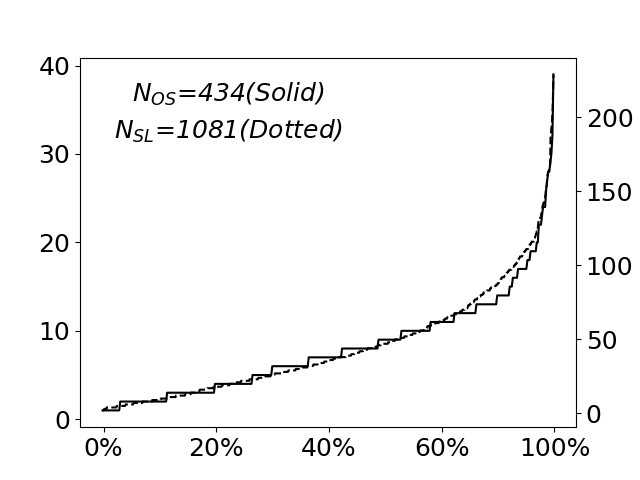}\hfill
       \includegraphics[width=.24\textwidth]{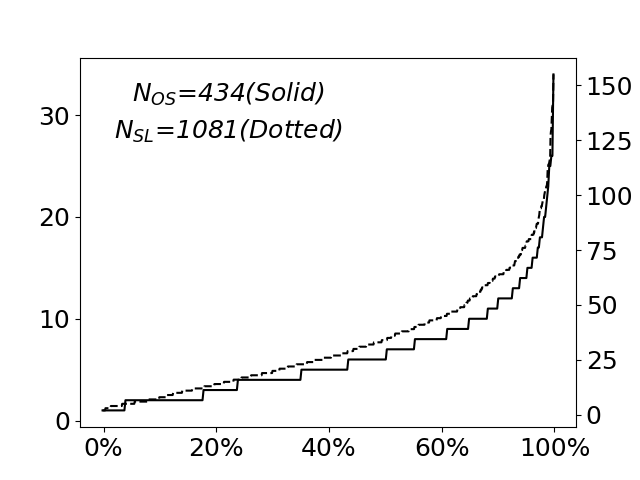}\hfill
     \includegraphics[width=.24\textwidth]{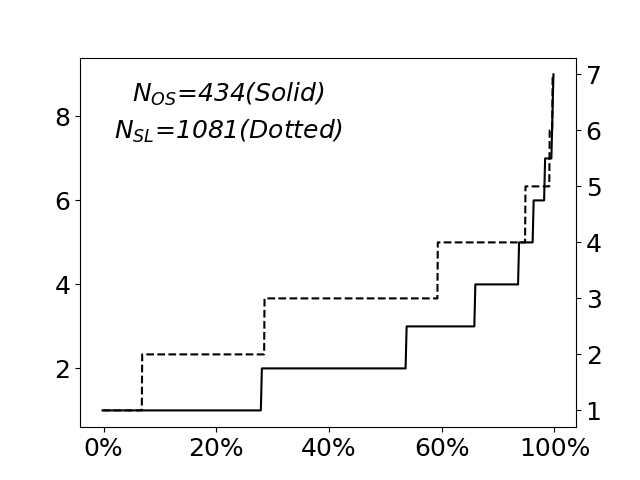}\hfill
   \includegraphics[width=.24\textwidth]{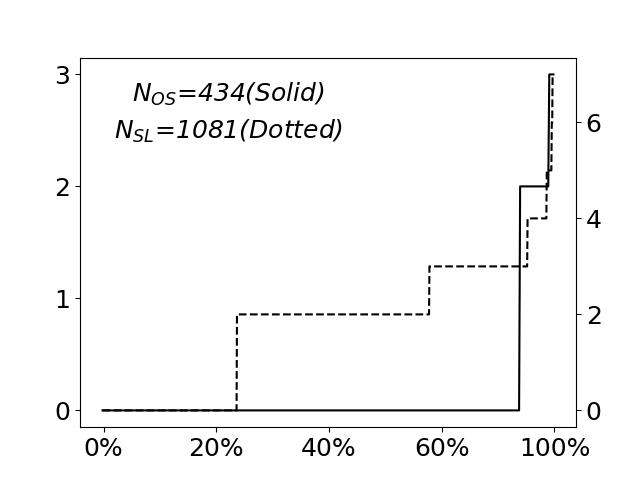}\hfill
    \\
\includegraphics[width=.24\textwidth]{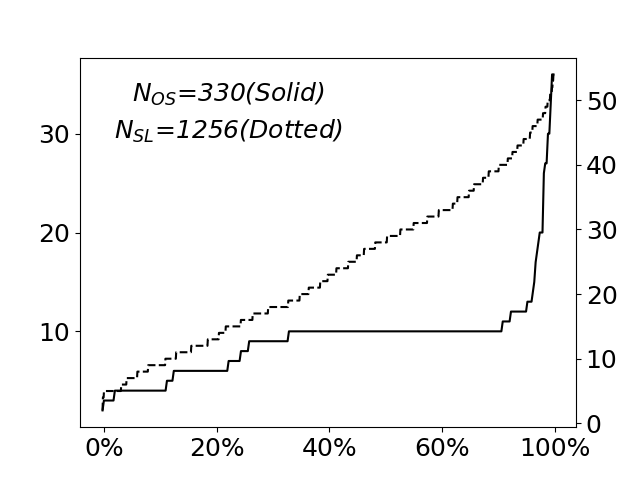}\hfill
   \includegraphics[width=.24\textwidth]{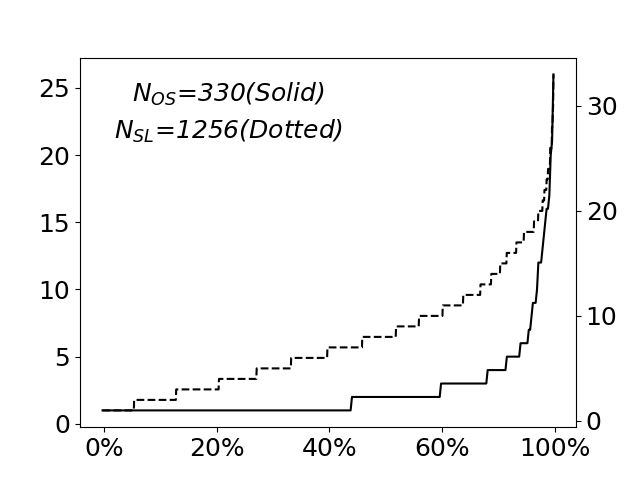}\hfill
    \includegraphics[width=.24\textwidth]{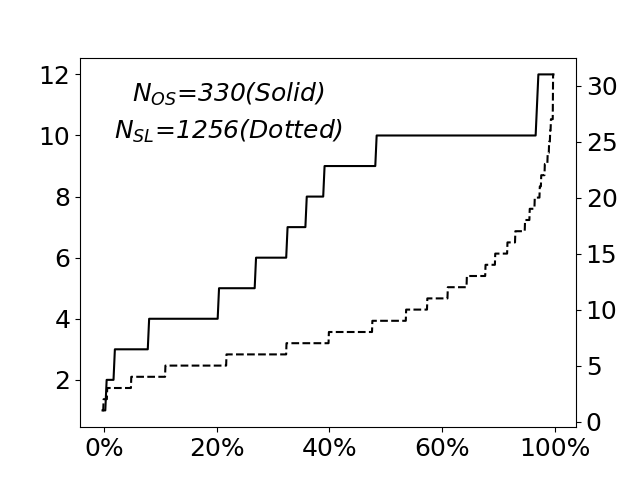}\hfill
   \includegraphics[width=.24\textwidth]{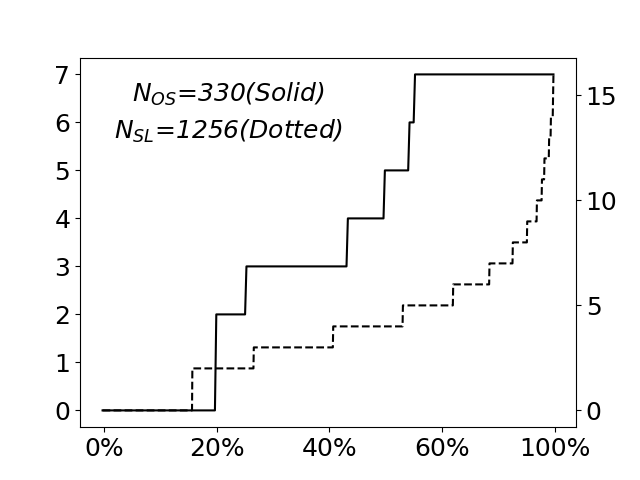}\hfill

\caption{Training models (top), DeepFuzzSL-generated models (middle), and \toolName{}-generated models (bottom). Left y-axis is for open-source training models and right y-axis is for SLforge-generated training models. X-axis is (valid) Simulink models sorted in ascending order for metric (from left to right column): Blocks per model, connected subgraphs, blocks in largest connected subgraph, maximum path length from a source to a sink block. Metrics of \toolName{}-generated models are overall closer to the training models than DeepFuzzSL-generated models.
}
\label{fig:distributiondomparison}

\end{figure*}

%On the flip side, 
The valid models generated by DeepFuzzSL were not as similar to the training models as \toolName{}-generated valid models. Figure~\ref{fig:distributiondomparison} compares these models along four metrics. For example, DeepFuzzSL-generated models tend to have many subgraphs that only contain 2 blocks, many blocks have unconnected input and output ports, and there is often no connection between source and sink.

\subsection{\toolName{} Found Superset of Bugs DeepFuzzSL Found (RQ2)}

Trained on SLforge-generated Simulink models, from nearly 3k \toolName{}-generated models 13 crashed Simulink. Upon analysis these 13 instances belong to the same two bug categories DeepFuzzSL found (MathWorks confirmed both types as known bugs). The first issue is a Simulink crash while opening a model. The second issue is Simulink opening a model but crashing while compiling the model. 

Trained on our open-source models, 
30 DeepFuzzSL-generated and 14 \toolName{}-generated models crashed Simulink while compiling. 
13 of the \toolName{}-generated (and all DeepFuzzSL-generated) models get rejected by Simulink R2018b but crash version R2020b (case 04777147). The last one crashes R2018b but is accepted by R2020b (case 04767975). Following are brief summaries of these two cases.

\subsubsection{Case 04777147 (Non-bug)} This \toolName{}-generated Simulink model triggered an interesting behavior, where Simulink R2018b rejects it as corrupt and the newer R2020b version crashes.
MathWorks told us that the way Simulink parses MDL files has changed since R2020a, which may have caused the crash. As it is impossible to create this model via Simulink's graphical editor or standard API, MathWorks Support marked it as a non-bug.
DeepFuzzSL-generated models triggered similar Simulink crashes.

\subsubsection{Case 04767975 (Known bug)}

\begin{figure}[ht]
    \centering
    \begin{minipage}{.2\textwidth}
        \centering
        \includegraphics[scale=\scaleFactor]{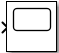}
    \end{minipage}
     \begin{minipage}{.2\textwidth}
        \centering
        \includegraphics[scale=\scaleFactor]{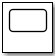}
    \end{minipage}
    \caption{Scope (left) and Floating Scope (right).}

    \label{fig:scope}
\end{figure}

Figure~\ref{fig:scope} shows two types of Simulink scope blocks: Scope and Floating Scope. Floating Scope does not have any physical ports while Scope does. A \toolName{}-generated model set floating parameter off (indicating that it is a normal scope) while setting the ports attribute to 0 (instead of a vector), causing the crash. Simulink's graphical editor or standard API cannot create this model.
This issue exists in R2018b and has been fixed in later versions. DeepFuzzSL did not trigger this bug.

\section{Related Work}

Small training datasets are a common problem in deep learning applications. Researchers thus often use synthetic datasets~\cite{SE:keepitsimple}.
Robbes et. al. showed a promising avenue to alleviate the dataset problem by using transfer learning~\cite{SE:pretrainedforSE}, i.e., that a small natural-language software engineering dataset can be used to improve sentiment analysis using pre-trained neural networks.

In the CPS domain, Chowdhury et al. developed a randomized differential testing tool using semi-formal specifications to test the Simulink toolchain~\cite{chowdhury2018icse}. Subsequently SLEMI generated semantic-preserving mutants of a seed model for differential testing of the Simulink toolchain~\cite{Chowdhury20SLEMI}. While these approaches are tightly coupled with Simulink,
\toolName{} is only loosely coupled and does not rely on explicit Simulink language specifications. 

Success of modeling natural language using deep learning has garnered interest to model source code for program generation. 
Researchers have used language models to improve software engineering task such as code completion and code clone detection~\cite{deepSE:codeCompletion,deepSE:codeclone,deepSE:codesummarization}. 
For compiler validation, DeepSmith~\cite{deepFuzz:Cummins}, DSmith~\cite{deepfuzz:Dsmith}, and DeepFuzz~\cite{deepfuzz:deepfuzz} uses deep learning based sequence modeling to model the OpenCL and C languages from real world programs and found compiler bugs. 
All of these approaches target languages with complete available specifications while we target Simulink, which does not have such a specification publicly available.  
 
The most closely related work DeepFuzzSL~\cite{Shrestha20DeepFuzzSL} use LSTM architecture to model Simulink. However they only train on synthetic models, citing the need for a larger training corpus. In contrast, we use a pre-trained language model and fine-tune it with open-source Simulink models.

Transfer learning for source code modeling is relatively new. Benito et al. studied the use of pre-trained models for source code generation and completion~\cite{TransferSE:CodeGeneration}. Hussain et al. proposed a transfer-learning based attention learner approach to improve code suggestions~\cite{TransferSE:DeepTransferLearning}. While earlier work focused on traditional languages we focus on a graphical CPS language.

\section{Conclusions}

Testing a commercial CPS development tool such as Simulink is hard as its codebase contains millions of lines of code and complete formal language specifications are not available.
While deep learning techniques promise to learn such language specifications from sample models, deep learning needs a large number of training data to work well. \toolName{} addressed this problem by using transfer learning, to leverage the powerful GPT-2 model that has been pre-trained on a large set of training data. \toolName{} adapted GPT-2 to Simulink with both randomly generated models and models mined from open-source repositories. \toolName{} produced Simulink models that are both more similar to open-source models than its closest competitor, DeepFuzzSL, and found a super-set of the Simulink development toolchain bugs
found by DeepFuzzSL.

\begin{acks}
The authors acknowledge the Texas Advanced Computing Center (TACC) at The University of Texas at Austin for providing HPC resources that have contributed to the research results reported within this paper.
Christoph Csallner has a potential research conflict of interest due to a financial interest with Microsoft and The Trade Desk. A management plan has been created to preserve objectivity in research in accordance with UTA policy. This material is based upon work supported by the National Science Foundation (NSF) under Grant No. 1911017 and a gift from MathWorks. 
\end{acks}

\bibliographystyle{ACM-Reference-Format}
\bibliography{ref}

\end{document}